\documentclass{emulateapj}

\usepackage{color}
\usepackage{graphicx}
\usepackage{natbib}

\begin{document}

\title{Seeking Counterparts to Advanced LIGO/Virgo Transients with Swift}

\author{Jonah Kanner, Jordan Camp}
\affil{NASA Goddard Space Flight Center, Mail Code 663, Greenbelt, MD 20771}
\email{jonah.b.kanner@nasa.gov}
\author{Judith Racusin, Neil Gehrels}
\affil{NASA Goddard Space Flight Center, Mail Code 661, Greenbelt, MD 20771}
\and
\author{Darren White}
\affil{University of Sheffield, Hicks Building, Hounsfield Road, Sheffield S3 7RH, United Kingdom}

\begin{abstract}
Binary neutron star (NS) mergers are among the most 
promising astrophysical sources of gravitational wave emission for
Advanced LIGO and Advanced Virgo, expected to be 
operational in 2015.
Finding electromagnetic counterparts to these signals
will be essential to placing them in an astronomical context.  
The \textit{\textit{Swift}} satellite carries a sensitive X-ray telescope (XRT),
and can respond to target-of-opportunity 
requests within 1-2 hours, and so is uniquely poised to find
the X-ray counterparts to LIGO/Virgo triggers.  
Assuming NS mergers are the progenitors of short gamma-ray bursts (GRBs),
some percentage of LIGO/Virgo triggers will be accompanied by X-ray band
afterglows that are brighter than 10$^{-12}$ ergs/s/cm$^2$ in 
the XRT band one day after the trigger time.  
We find that a soft X-ray transient of this 
flux is bright enough to be extremely rare,
and so could be confidently associated with even a moderately
localized GW signal.  
We examine two possible search strategies with the 
\textit{Swift} XRT to find bright transients in LIGO/Virgo
error boxes.
In the first strategy, XRT could search a 
volume of space with a $\sim$100 Mpc radius by observing $\sim$30 galaxies 
over the course of a day, with sufficient depth to
observe the expected X-ray afterglow.  For an extended LIGO/Virgo
horizon distance, the XRT could employ 100 s 
exposures to cover an area of $\sim$ 35 square degrees in about
a day, remain sensitive enough to image
GW discovered GRB afterglows.  These strategies demonstrate that
discovery of X-ray band counterparts to GW triggers will be possible.

\end{abstract}


\maketitle

\section{Introduction}

The construction of the Advanced LIGO and Advanced
Virgo 
\footnote{https://tds.ego-gw.it/itf/tds/file.php?callFile=VIR-0027A-09.pdf}
gravitational wave (GW) observatories 
is in progress, with completion expected as early 
as 2015 
\citep{advLigo,ivirgo}.  
These detectors are expected to serve
as all-sky monitors for mergers of binary 
neutron stars (NS-NS), mergers of neutron stars with 
stellar mass black holes (NS-BH), and mergers of 
binary black holes.  
The detectors are designed to observe NS-NS mergers
to an average distance of 200 Mpc, and
NS-BH mergers to 400 Mpc 
 \citep{rates}.  The GW signal from such
events could provide a wealth of information, including
the masses and spins of the component objects, and an 
estimate of the luminosity distance to the source.  
However, this network of gravitational wave detectors
will have, even in the best case scenarios, only modest 
localization ability.
This means that placing a GW observation in an 
astronomical context, including identification of a 
host galaxy and environment, will require finding a
counterpart electromagnetic (EM) signal to the merger
event.  A joint EM/GW observation would describe an
explosive event in unprecedented detail, since the GW 
signal would directly probe the progenitor's dynamics
while the EM signal would carry information on the environment 
and allow improved parameter estimation.
Moreover, a population of such joint EM/GW signals could 
be used to estimate cosmological parameters \citep{sirens, nissanke}.  

The main challenge in the identification of an EM counterpart
to an observed GW signal will be the large positional uncertainty 
associated with current networks of GW observatories.
The positional uncertainty for a given event will depend strongly
on a number of factors, including signal-to-noise ratio (SNR),
position on the sky, available 
position reconstruction algorithms, and the internal state of the 
GW network.  
As designed, all three detectors in the 
Advanced LIGO/Virgo network would operate at similar sensitivity, 
leading to typical positional uncertainties around 
$\sim 20$ square degrees  
\citep{cwbposrec, Fairhurst2011, nissanke2011}.
On the other hand, studies using data from the last science 
runs of LIGO and Virgo have shown uncertainties for low 
SNR signals of 50-200 square degrees \citep{s6methods},
during times when detectors differed
by a factor of $\sim 2$ in amplitude sensitivity.
During the early years of observing with the second 
generation network, evolving sensitivity levels
will likely lead to variation in the ability to localize 
sources.  

Since there have been no certain observations of
stellar mass compact object mergers,
predicting the wavelength, flux, and duration 
of a possible EM signal is somewhat speculative.  
However, some models seem promising. 
NS-NS mergers and NS-BH mergers are both possible
progenitors for short gamma-ray bursts (GRBs) \citep{foxNature, gehrelsnature},
and both their prompt emission and afterglow
have been carefully studied.  In addition, 
theoretical  considerations and simulations lead to the 
expectation of an isotropic, optical signal
that results from energy released in decays
of unstable isotopes following r-process nucleosynthesis
\citep{metzger, enrico, piran2012}.  
Some consideration has already been given to
observing strategies that might
discover an optical or radio band counterpart
to a merger event \citep{metzberg, coward_optical}, and searches have 
been performed in a range of wavelengths on low-threshold
triggers from the initial LIGO/Virgo network
\citep{swiftfollow, s6methods, cbcquick}. 
Past searches have also sought coincident triggers using
archived GRB and GW data \citep{s6grb}.
In this work, we focus on the
X-ray band, and consider the \textit{Swift} satellite \citep{swift}
as a potential 
instrument for discovering EM signals following 
a trigger from the second generation LIGO/Virgo network.    

\textit{Swift} has an unmatched record as 
an engine for producing observations
of GRBs and their afterglows.  In the typical 
mode of discovery, \textit{Swift}'s Burst Alert Telescope \citep{bat}
sweeps the sky for GRBs.  When a GRB is discovered,
the X-ray Telescope (XRT) \citep{xrt} and UV/Optical Telescope
(UVOT) \citep{uvot} are automatically slewed to the estimated source
position.  This strategy has been extremely successful:
the XRT finds soft-band X-ray counterparts to 
nearly $80\%$ of observed short GRBs with prompt observations.  
For comparison, optical band counterparts are only discovered for 
$\sim30\%$ of short GRBs.  If we accept the 
putative link between compact object mergers and 
short GRBs, then the XRT band (0.3-10 keV) seems a natural
place to find the counterparts to compact object 
mergers.

\section{Sources}

\subsection{GRB Afterglows}

\begin{figure}
\begin{center}
\includegraphics[width=0.95\columnwidth]{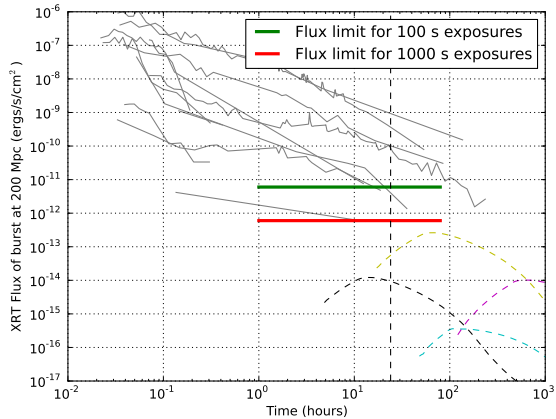}
\caption{The gray curves are XRT light curves for short GRBs 
with known redshifts, scaled to a distance of 200 Mpc
\citep{xrt_repo1, xrt_repo2}. 
The green line shows XRT flux limit for 100 s exposures,
which could cover $\sim35$ square degrees in one day,
and the green line indicates the limit for a 1000 s exposure.
Some observed afterglows quickly fade, and so are not observable 
1 hour after the burst.
However, the ``long-lived'' afterglows are generally bright enough
to be observed 10-100 hours after the burst, even with short ($\sim$100 s) exposures.  The dashed, colored curves show predicted light curves
for off-axis light curves viewed at twice the jet opening angle, 
scaled to 200 Mpc \citep{eerten2}.  The simulated light curves have jet energies
of $10^{48}$ and $10^{50}$ ergs, and circumburst medium number 
densities of 1 cm$^{-3}$ and $10^{-3}$ cm$^{-3}$.
}
\label{xrt_light}
\end{center}
\end{figure}

The collection of short GRB afterglows
observed with \textit{Swift} and with measured
redshifts has been studied by 
\citet{judy2011}.  The late-time X-ray afterglows
decay with a power law $t^{-\alpha}$, with a temporal
index $\alpha \sim 1.5$
and at 1 day after the trigger
time, show luminosities in the XRT band 
ranging from $10^{42}-10^{45}$ ergs/s.  
Placing these afterglows at a luminosity distance
of 200 Mpc leads to fluxes of between 
$10^{-12}$ and $10^{-9}$ 
ergs/s/cm$^2$ (See Figure \ref{xrt_light}).  The XRT routinely observes
these objects out to a redshift of $z \sim 0.5$,
suggesting that the afterglows from sources within
the GW detector horizon ($\sim 400$) Mpc would be relatively bright.

An important feature of afterglows in this context
is that they are expected to be beamed.  This 
has the implication that only a small fraction, $f_b$, of 
NS-NS or NS-BH mergers will have a beam pointed towards Earth.
For small jet opening angles, $f_b$ is related to the jet
opening half-angle as $f_b \sim \theta_j^2/2$.
The jet angle is highly uncertain, but is typically
expected to be between a few degrees and a few tens of 
degrees, meaning that $f_b$ is likely of order 
$1\%$.  If this is the case, then there is a 
major implication for LIGO/Virgo triggers: the vast majority
of them will not be associated with on-axis GRBs.  
However, the fraction of GW selected events
within the beam will likely be larger than $f_b$ due to 
a particular bias \citep{bernie2011, nissanke}.
Inspiraling compact objects emit gravitational 
wave energy preferentially in the direction parallel
to their angular momentum axis, that is, 
the gravitational waves are weakly beamed in the 
same direction as the GRB jet.  
\citet{bernie2011} shows that 
this effect will increase the fraction of 
GW selected compact object mergers with their 
beam pointed towards earth by a factor of 
3.4 over the strictly geometric prediction.
Moreover, a few short bursts have lower limits 
on their jet opening angles placed above $10$
degrees \citep{judy2011, wenfai, coward_grb}.  
If we estimate the opening angle 
of short GRBs as around 10 degrees,
then the expected fraction of LIGO/Virgo 
observed mergers with earth within the 
jet is $\sim 5 \%$, or one observable GRB in every
$\sim 20$ LIGO/Virgo observed NS-NS mergers.

The above estimate of $f_b$ is highly uncertain,
due to the limited number of observations of
short GRB jet breaks.  On the other hand, it is
possible to make a reasonably robust estimate of the
number of short GRBs within the LIGO/Virgo range,
based on the observed rates of GRBs
\citep{holz2012, metzberg, s5grb}.  The main source
of uncertainty is then the progenitor of short bursts.
If we assume that all short GRBs are due to NS/NS mergers,
then we expect $0.3 - 3$ events in range per year with
Advanced LIGO design sensitivity.  However, the LIGO
range for BH/NS events is greater by a factor of $\sim 2$,
leading to $2 - 20$ events per year if we assume all 
short GRBs are caused by NS/BH mergers.  In the early years
of Advanced LIGO, with half the design range, 
we might then expect $\sim 0.1$ 
GW/GRB coincidences per year under the NS/NS assumption,
or $\sim 1$ such event per year under the BH/NS assumption.
Uncertainties include
the possible
gain in reach of the GW instruments due to reduced background
with a coincident GRB observation, or factors accounting for 
the less-than-perfect spatial and temporal coverage of 
both GRB and GW monitors.  

Even if a short GRB 
is beamed in a direction away from the earth,
X-ray band emission from the afterglow may be
visible in the direction of earth.  There are 
no confirmed observations of such off-axis afterglows, presumably
due to the difficulty in reliably identifying them
in all-sky survey data.  However, some predicted light 
curves from off-axis afterglows are available
from simulations by \citet{eerten2}.  
In order to compare the simulation results with 
observed, on-axis light curves, we scale the results
to a luminosity distance of 200 Mpc, and calculate
the XRT band flux assuming a power law spectrum with an
index of 1.5 (see Figure \ref{xrt_light}).  
Comparing off-axis light curves with observations
of GRBs (where the observer is inside the jet opening
angle) suggests that the off-axis observer will encounter
a number of challenges seeking an X-ray counterpart.  For an observer located
at twice the jet opening angle, there is a brightening
time of between two and twenty days, so 
a wait of at least several days would be needed to observe
an off-axis afterglow in X-rays.  A large 
time window between the GW trigger and the observation 
of the counterpart would make establishing a connection
difficult.
Moreover, the off-axis emission is considerably dimmer than the 
on-axis emission, by several orders of magnitude (See Figure 
\ref{xrt_lim_flux}).
In addition to being more difficult to detect, the fainter
emission will also be more difficult to separate from
background variability.  This suggests that an
observation of an off-axis afterglow would require
an exceptionally nearby NS-NS merger.  For a merger
at 50 Mpc, the peak flux of an X-ray afterglow observed at 
twice the jet opening angle would 
range $\sim 10^{-15} - 10^{-13}$ ergs/s/cm$^2$, and so 
could be observable in a 10 ks XRT exposure.  Such events
are likely to be rare, perhaps once every 10-20 years based
on observed GRB rates.  However, they present 
an interesting possibility, since the large SNR signal
that they would produce in the Advanced LIGO/Virgo network
would allow for exceptional localization and parameter estimation.  

\subsection{Kilonovae}

While observing the afterglow to a short GRB from a 
position outside the beam will likely be challenging, 
another source of emission from NS-NS mergers may create
an isotropic transient.  Though lacking in 
observational evidence, transients known as kilonovae have 
been treated in the literature by several authors
\citep{metzger, enrico, lipac, macro, goriely}.  The model predicts that
ejecta from the merger will grow heavy nuclei through
r-process nucleosynthesis, which subsequently decay and 
heat the material.  The thermal emission leads to an 
optical band transient with a peak luminosity around 
one day after the merger event, and a dimming over the course
of the next few days.  The transient may have a blue color, 
with $U$ band emission that appears brighter and peaks sooner 
than the $R$ band emission.  Because this mechanism
is largely independent of the environment around the merger,
and leads to isotropic emission, it is possible to imagine
that a large fraction of NS-NS and NS-BH mergers are 
accompanied by observable kilonovae.
A kilonova at 200 Mpc is expected to peak around magnitude
19-22, bright enough to be 
detected by the UVOT instrument on \textit{Swift}.  UVOT 
is aligned in parallel to the XRT,
so a search over one or two days 
for X-ray afterglows with \textit{Swift} is a simultaneous
search for optical band kilonovae.
It should be noted that Swift's capability in searching 
for optical band transients is not unique; ground-based
optical survey telescopes can also search large areas
to these magnitudes \citep{metzberg}.

\section{Search Strategies with \textit{Swift}}

The large position uncertainty associated with
LIGO/Virgo triggers matches well onto the capabilities 
of ground based, large area survey projects such as 
PTF, Pan-STARRS, QUEST, and SkyMapper \citep{metzberg}. 
However, a space-based
X-ray facility may present some unique advantages. 
The X-ray afterglows to short GRBs
are easily distinguished from other X-ray sources both by 
their large flux and characteristic power-law dimming.  
A space-based, as compared to ground-based, facility also
has the advantage that wait time for a source to pass overhead
is typically $\sim 90$ minutes instead of $\sim 12$ hours, 
and the sky coverage is nearly total, where a ground based 
facility has access to a smaller fraction of the sky.  

The combination of
fast response times to target-of-opportunity (TOO) 
requests (around 1 or 2 hours),
and a long heritage with GRB afterglows
makes the \textit{Swift} satellite a natural facility to consider.  
Much of the following discussion could apply to other
facilities as well, particularly XMM-Newton and Chandra, 
however the fast TOO response may make \textit{Swift} the only practical
choice for seeking quickly fading counterparts.  With 
this in mind, and with the intention of making the 
discussion as concrete as possible, we focus specifically
on the \textit{Swift} observatory.  

\subsection{Searching the full error box}

The first strategy that we consider is using
the \textit{Swift} XRT to tile an entire LIGO/Virgo error box.  
Estimates for the uncertainty associated with a LIGO/Virgo
position reconstruction vary from a few tens of square degrees
to over a hundred square degrees 
\citep{s6methods, nissanke2011, cwbposrec, Fairhurst2011}.  
In fact, the precision of
any particular position estimate will depend on a number of factors,
including SNR and sky position.  Certainly, any position reconstruction
with the LIGO/Virgo network will cover an area significantly larger than the 
0.16 square degree XRT field of view.  In order to evaluate 
the feasibility of search strategies, we consider 100 square degrees
as a nominal value for the LIGO/Virgo position uncertainty.

\begin{figure}
\begin{center}
\includegraphics[width=0.95\columnwidth]{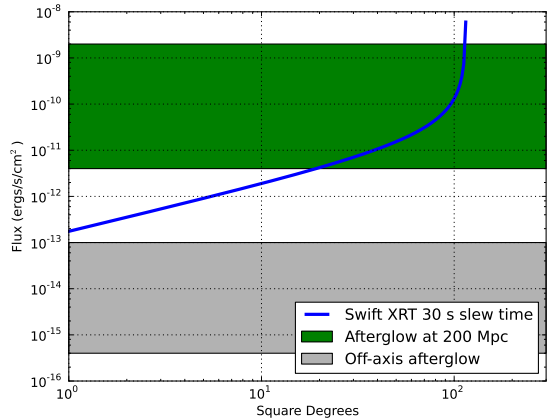}
\caption{ Plot showing the potential for a large area survey with 
XRT over the course of a single day.   Enough exposures are 
taken over the course of 24 hours to cover the area 
shown on the 
horizontal axis, under the assumption that the XRT requires
30 s of slew and settle time for each image. 
For comparison, the range of fluxes of short GRB afterglows,
scaled to 1 day after the trigger time and a luminosity distance
of 200 Mpc, is shown as the green shaded region.  The 
gray region shows the peak XRT flux of a range of off-axis
lightcurves, also scaled to 200 Mpc.  The peak flux
of the model off-axis light curves occurs several days
after the GRB trigger time.  
}
\label{xrt_lim_flux}
\end{center}
\end{figure}

To characterize the ability of the XRT to quickly
survey a large area, we write the limiting flux
of an observation as 
a function of observing time as
\begin{equation}
F = 6 \times 10^{-12} \left( \frac{100~\textrm{s} }{T} \right) \textrm{ergs/s/cm}^2
\end{equation}
where $T$ is the observing time \citep{moretti}.  This is valid 
in the regime of photon limited exposures ($T < 10^4$ s),
and assumes 12 counts are needed for a detection, with a 
conversion factor between count rate and flux of $5\times 
10^{-11}$.     
If we imagine that
the sought afterglow to a GW trigger is visible for 
roughly one day, then we can calculate how much observing
time, and so what limiting flux, we can associate with 
observations covering various amounts of area.
\begin{equation}
T = \left( \frac{86,400 ~ \textrm{s}}{3} \right)  \left( \frac{0.12 ~\textrm{deg}^2}{\textrm{Area}} \right) - S
\end{equation}
\begin{equation}
F \approx 2 \times 10^{-14} \left( \frac{\textrm{Area}}{0.12 ~\textrm{deg}^2} \right) \textrm{erg/s/cm}^2
\label{flux_eqn}
\end{equation} 
where $S$ represents the amount of time for each slew and settle
of the instrument.  Equation \ref{flux_eqn} assumes the observing time
is much larger than the total slew time.  
The factor of $\frac{1}{3}$ is to account
for the fact that most observations cannot continue over an entire
orbit, due primarily to occultation by the earth.  
In addition, the size of the field of view has been reduced to 0.12 deg$^2$
to allow some overlap in the tiling pattern.  
The resulting limiting fluxes are plotted in 
Figure \ref{xrt_lim_flux} for a range of areas, with 
$S$ set to 30 s.

Under these assumptions,
the XRT might take a series of $290$ exposures, 
each 100 s long.  This would cover 35 deg$^2$ in about
one day, and yield a flux limit of around $6 \times 10^{-12}$ ergs/s/cm$^2$.
With the current on-board software, this could be accomplished
with eight applications of the programmed 37 tile pattern.
The resulting flux limit seems to be
sensitive enough to find most on-axis afterglows, but
not predicted off-axis afterglows for most viewing angles.

Clearly, this approach would be less strenuous if the position uncertainty 
associated with a particular event could be reduced.  
Some estimates do predict that 100 square degrees is a conservative
estimate, and suggest 20-40 square degrees as more typical 
numbers, depending on underlying signal models and algorithm.  
For example, \citet{nissanke2011} predict
that, using a Markov Chain Monte
Carlo parameter estimation technique, only 36 square degrees
of area need to be searched to recover $70\%$ of binary 
mergers detected with an SNR of at least six 
at all three detector locations, or 12 square degrees 
could be searched for a $50\%$ recovery rate.
The addition of a fourth GW detector 
to the network could also improve localization ability to 
around $10$ square degrees.  KAGRA \citep{kagra}
located in the Kamioka mine in 
Japan, 
could be operational by 2018, and a third LIGO site in 
India \footnote{https://dcc.ligo.org/cgi-bin/DocDB/ShowDocument?docid=91470}
could be operational by 2022.

\begin{figure}
\begin{center}
\includegraphics[width=0.95\columnwidth]{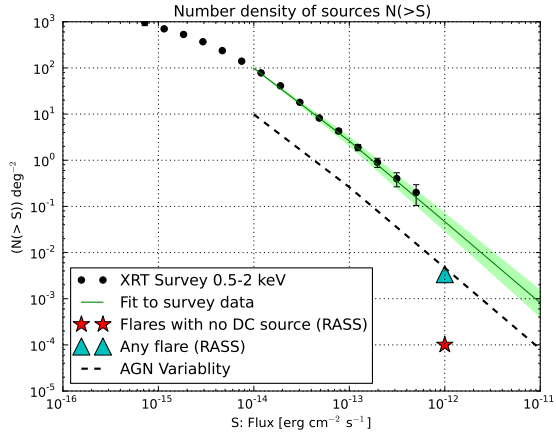}
\caption{ The extragalactic X-ray background number density
estimated in various ways.  The black points show
the observed number density at high galactic latitude 
in the \textit{Swift} serendipitous survey, with a fit shown in green \citep{swiftss}.
The black dashed line is an estimate of 
the number of variable AGN that may be mistaken as transients, 
assuming a reference image that is twice the depth of the 
limit shown on the x-axis \citet{agnvariable}.  The triangle marks the density
of variable ``flare like'' sources in the ROSAT All-sky survey \citep{rassvari},
and the red star marks the density of candidate orphan afterglows
found in a ROSAT data \citep{rassgrb}.  
}
\label{xrt_back}
\end{center}
\end{figure}

An important consideration in searching a large area for
a single event is to understand if we can distinguish the 
event from other sources.  For the case where the 
error box is out of the galactic plane, some suggestive
numbers are shown in Figure \ref{xrt_back}.  
At a sensitivity of $2 \times 10^{-12}$ erg/s/cm$^2$,
a typical area of 100 square degrees would find
only a few extragalactic sources \citep{swiftss, mateos2008}.  Demanding variability
in the source should be a strong handle for cutting this number
down further.  The black dashed curve uses statistics
of observed AGN variability to forecast how many variable
AGN are likely to fluctuate by a factor of 2 or more 
between two images (based on numbers found in \citet{agnvariable}).  
The study represented as a 
blue triangle found a similar number density by 
systematically searching the RASS data for variable sources with a 
``flare'' like light-curve \citep{rassvari}.  The red star
shows the result of another RASS study that used stronger cuts
to seek GRB orphan afterglows \citep{rassgrb}.  The three orders of 
magnitude in density reduction between this point and the 
number of AGN was achieved by demanding that 
no X-ray source was present at the location of the transient 
either before or after the flare event.  This seemed to 
be a very powerful cut, and resulted in finding around
one source in every 10,000 square degrees, or 
a $1\%$ chance of finding an unrelated afterglow like 
source in association with the 100 square degree 
GW error box.  When seeking only afterglows within 
a limited distance range, it is possible to remove the 
majority of flare stars by demanding a spatial coincidence
between the transient source and an optical galaxy.
This suggests that a transient, soft
X-ray source, brighter than $\sim 10^{-12}$ ergs/s/cm$^2$
found in connection to a GW trigger is likely to be 
associated.  The expectation that some LIGO/Virgo triggers
will have short-lived, soft X-ray counterparts brighter than anything 
else within the error box suggests that a wide-field,
focusing instrument in this band would be a useful follow-up tool.

A LIGO/Virgo trigger in temporal coincidence with a 
\textit{Fermi} GBM \citep{gbm} trigger would be of great interest. 
GBM sees a large fraction of the sky ($\sim 65\%$) \citep{gbm}, and 
so this is a likely scenario.  The GBM has a large
uncertainty in its position reconstruction \citep{gbmposition}.
For example, in the GBM burst catalog \citep{burstcat},
the listed short bursts have a median positional uncertainty
of 7.5 degrees.  Some of this 
error ($\sim 3$ degrees) is due to systematics 
in the localization process \citep{gbmipn, burstcat}.

\begin{figure}
\begin{center}
\includegraphics[width=0.95\columnwidth]{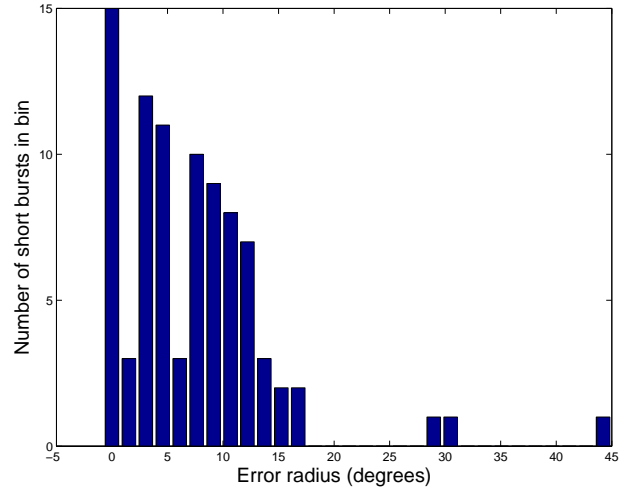}
\caption{A distribution of estimated error radii for 88 
bursts from the \textit{Fermi} GBM Burst Catalog.  The bursts
are selected only with the criteria that T$_{90} < 2$ s.
The x-axis is in degrees, the y-axis is the number of bursts
in the bin.  Many of the bursts in the $0$ bin 
have position 
information that is known from some other source, such as the 
\textit{Fermi} Large Area Telescope or \textit{Swift}.
}
\label{fermirad}
\end{center}
\end{figure}

A GW trigger in coincidence with 
a \textit{Fermi} GRB observation would definitively establish the progenitor
of the burst.  However, the large error radius associated
with \textit{Fermi} means that a GBM trigger alone could not
provide a host galaxy identification or provide 
the coordinates of the potential afterglow.  Given
the high degree of interest in such an event, 
seeking the afterglow and host galaxy seems well worth
the effort.  The range of typical GBM position uncertainties
(3 - 12 degrees, see Figure \ref{fermirad}) 
overlaps the range of positional errors
with LIGO/Virgo, with a median position 
uncertainty of 7.5 degrees, or 176 deg$^2$
under the assumption of a circular region.  
While these uncertainties may often be 
larger than the GW position uncertainties, in the early
days of advanced detectors, it is possible that
the three GW detectors will have unequal sensitivities.  
This would lead to GW error boxes that are very spread
out on the sky.
For example, if only two detectors are operating at the 
time of an event, then the GW network will localize
the event to a ring on the sky of order 1000 square degrees.  
In such a scenario, the \textit{Fermi} error ellipse would be 
more constraining the LIGO/Virgo uncertainty region.  
It is also possible
to imagine taking the intersection of the GBM error ellipse
and the LIGO/Virgo skymap.

\subsection{Searching with a galaxy catalog}

The fact that the LIGO/Virgo network is sensitive to
only a limited distance can be used to dramatically
reduce the amount of area that must be searched for an
EM counterpart.  Merger events should occur within,
or close to, their host galaxies.  In fact, 
\citet{bergerenv} has found that short GRBs seem to 
track the total mass of galaxies.
So, in response to a LIGO/Virgo trigger, 
it may be possible to search only the locations 
of known galaxies within a fixed distance horizon
rather than the entire error box.  
This method was applied in efforts to search for EM afterglows to 
GW triggers using the initial LIGO/Virgo network \citep{kanner2008, s6methods}.
For the 2009-2010 search, a galaxy catalog was constructed 
from publicly available information,
known as the GW Galaxy Catalog (GWGC)\citep{gwgc}.  
This approach takes advantage of the limited distance reach of the 
GW instruments to merger events, in the sense that the sky
is relatively sparse in galaxies to a limited distance range.
In the limit of a 
GW detector that could observe events anywhere in the observable
universe, this would clearly not be the case, and the 
density of observable galaxies on the sky would make a galaxy catalog
ineffective in limiting the amount of sky area to be observed.  
The question that we ask is:
out to what distance reach is a galaxy catalog still useful in
this way with the \textit{Swift} XRT?

A similar question was addressed by \citet{nuttall}, who 
found that the galaxy catalog can be an extremely useful tool in
recovering the true location of a GW inspiral signal out to 
at least 100 Mpc, even if only a few galaxies are 
imaged.  Here, we attempt to find the limiting
range where searches with and without a galaxy catalog
require the XRT to observe the same amount of sky area.

Past counterpart searches using a galaxy catalog assumed that the likelihood
of a merger event in the galaxy traces either the mass or 
the star formation rate of the galaxy \citep{s6methods, cbcquick}. 
Competing 
models exist for which types of galaxies are more likely 
to host merger events \citep{richard_whichgal}.  
In order to explore the implications of various models,
we have used the HyperLeda database to add measurements
of I-band (near infrared) luminosity to the GWGC.  
This resulted in a catalog
with 51,136 objects with $B$-band measurements, 
34,363 objects with $I$-band measurements, and 31,732 galaxies 
with both $I$-band and $B$-band measurements.
The $B$-band luminosity is estimated to be $\sim 60\%$ complete
\citep{gwgc}.  
Conventional wisdom suggests that wavelengths in the near 
infrared should be good tracers of total stellar mass.  
However, the application of a color correction has been shown
to improve mass estimates in some cases \citep{bell2001, bell2003}.

Adopting the model presented in \citet{bell2001}, we constructed a 
color corrected mass estimate using the $I$-band magnitude and 
$[B-I]$ color for each galaxy in our sample with both measurements.
We applied the model as
\begin{equation}
\log_{10} (M/L) = -0.88 + 0.60[B-I] 
\end{equation}
where $M$ and $L$ are the galactic mass and $I$-band luminosity,
both in solar units.  \citet{bell2003} showed that, in near
infrared wavelengths, this color dependency is likely too steep,
but did not provide corrected values for I-band measurements.  
For this reason, it is likely that our color corrected 
mass estimate amplifies the effects of color scatter due to 
metallicity variations between galaxies, and so results in an 
artificial broadening in the distribution of masses in the catalog.  
The effect was partially mitigated by removing galaxies with colors
far from the center of the distribution.  

\begin{figure}
\begin{center}
\includegraphics[width=0.95\columnwidth]{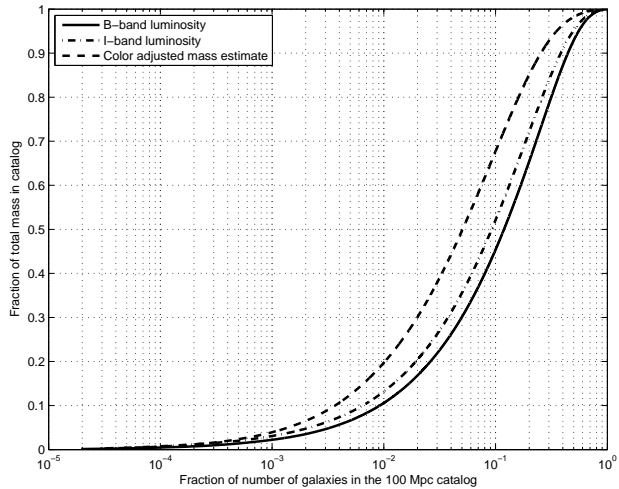}
\caption{ The GWGC contains roughly 53,000 galaxies within 100 Mpc of earth.
The plot shows what fraction of the total number of galaxies must be selected
in order to obtain a target fraction of the total luminosity or mass
in the catalog.  Only 10\% of the galaxies contain 50\% of the I-band
luminosity.  Including 90\% of the blue light luminosity requires 40\%
of the number of galaxies.  The distribution of the color adjusted
mass estimate suggests that even a smaller fraction of galaxies
in the catalog may contain a given fraction of the total mass, 
however, this may be an effect of color scatter due to metallicity.
}
\label{gwgc_frac}
\end{center}
\end{figure}

\begin{figure}
\begin{center}
\includegraphics[width=0.95\columnwidth]{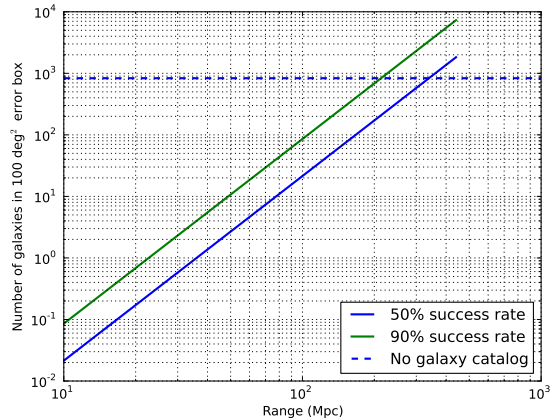}
\caption{ The GWGC contains 53,000 galaxies within 100 Mpc of earth.
The figure assumes that the number of galaxies within a horizon distance
$r$ scales as $r^3$, and that the catalog
is 60\% complete.  Within a 100 Mpc horizon distance, imaging the galaxies
needed to contain 90\% of the $I$-band luminosity within a 
100 square degree LIGO/Virgo
error box requires an order of magnitude less pointings of
the XRT than the number required to tile the whole LIGO/Virgo
error box.  The limit where a galaxy catalog is 
no longer useful seems to be around 200 Mpc.  For comparison,
the design curve for Advanced LIGO predicts a sky average NS-NS 
average range of around 200 Mpc.  Initial LIGO had a sky average
range of around 20 Mpc for NS-NS mergers.  
}
\label{gwgc_horizon}
\end{center}
\end{figure}

In the GWGC,
out to 100 Mpc there is roughly
1.3 possible host galaxies per square degree, or 130 possible hosts
for a typical LIGO/Virgo error box.  However, if we take the position 
that only enough possible hosts need to be imaged to make including the
true host likely, we can reduce this number further.  Figure \ref{gwgc_frac} shows
that, averaged over the sky, $40\%$ of galaxies contain 90 \% of the 
$I$-band luminosity.  After adjusting for the 
completeness fraction of the catalog, this means a
100 deg$^2$ LIGO/Virgo error box with a 
100 Mpc
range could be covered at the 90\% level by observing $\sim 90$ galaxies,
or at the 50\% level by observing $\sim 20$ galaxies.
This represents dramatically fewer pointings of the \textit{Swift} XRT than 
the 800 fields required to tile the whole error box.  So,
for a search with a range of 100 Mpc,
the galaxy catalog still seems to be a useful tool.  
This suggests that,
in the early months of Advanced LIGO/Virgo, when horizon distances are 
expected to be below the design goals, 
a reasonable strategy would utilize
exposures of roughly 1 ks, allowing the XRT to image 
2 galaxies per orbit, and so observe 32 galaxies in a 
24 hour period.  Repeating the observations over a 2nd,
not necessarily concurrent, day
would allow image subtraction with the UVOT to enable a search
for kilonovae.  
However, it is important to 
note that this scenario is limited to the assumption of a NS-NS merger:
BH-NS mergers have horizon distances that are greater by roughly a
factor of two.  Moreover, it is important to distinguish between the 
sky-averaged range distance, which we have used here, and the optimal
horizon distance, which differ by a factor of 2.26.

In the case of a very nearby source ($< 100 Mpc$), it may be possible
to limit observations to only a few galaxies.  In such cases, the 
search might be optimized for off-axis afterglows by taking longer,
$\sim 10 ks$ exposures of each galaxy, over a span of a few days.  
This would result in flux limits of around $2 \times 10^{-14}$ ergs/s/cm$^2$,
and so be deep enough for the more optimistic off-axis afterglow models.  

No galaxy catalog is currently complete to 200 Mpc, 
although there are efforts underway to obtain one by the time
of Advanced LIGO \citep{mansi_pri}.  Using the GWGC, a geometric scaling 
allows us to anticipate the result for a 200 Mpc
horizon distance (See Figure \ref{gwgc_horizon}.  At 200 Mpc, if we still 
hope to capture the true source with a 90\% likelihood,
and assuming a similar distribution of galactic luminosities
as those already contained in the GWGC, then the requirement is 
to observe around 700 galaxies per LIGO/Virgo trigger.  This number
is similar in scale to the 800 fields required to tile the whole 
error box, so we conclude that the limit of usefulness of a
galaxy catalog for 24 arcminute fields
is around 200 Mpc (See Figure 
\ref{gwgc_horizon}).  

Another important concern related to the application of a galaxy catalog
is the possibility that NS-NS or NS-BH mergers will occur {\it outside} of
their host galaxies.  Asymmetries in the supernova that 
forms a compact object can impart a net momentum, or ``natal kick'',
to the resulting NS or BH.  If the kick is large enough to 
unbind the system from the gravitational potential of its host galaxy,
then the binary can drift outside the host galaxy before merger.  
In fact, \citet{kalogera_kicks} found evidence for kicks around
200 km/s in observed neutron star binaries with separation distances
small enough to allow mergers within a Hubble time.  Considering
where mergers may be observed in relation to their host galaxies, 
\citet{kelley_kicks} found that if even larger kicks are assumed, 
$\sim 360$ km/s, then nearby mergers will be observed up to 1 Mpc 
away from their host.  However, other investigators find these
large kicks to be unlikely, and tend to favor models where 
typical mergers occur within 1 to 100 kpc of the host galaxy
\citep{kalogera_kicks, brandt_kicks, bloom_kicks, fryer_99, belczynski_kicks}.
Observational evidence also supports the notion that
most mergers occur within 100 kpc of the host galaxy.  An attempt
to match ``hostless'' short GRBs with nearby galaxies found that,
most of the observed short GRBs with 
known redshift very likely occurred within 100 kpc of their host galaxy
\citep{berger_nohost}.  In the same work, a comparison between the 
distribution of these
observed short GRB offsets from their host galaxies was found to be
consistent with predictions from models of NS-NS merger locations.  

The \textit{Swift} XRT has a FOV of 24 $\times$ 24 arcminutes, and the UVOT has a FOV of 
$17 \times 17$ arcminutes.  If a merger occurs at a distance of 100 Mpc from earth,
then an XRT (UVOT) observation centered on the host galaxy will 
observe the counterpart for any source within 300 kpc (200 kpc) of the 
host galaxy.  This should include essentially all NS-NS mergers.  Even
as close to earth as 50 Mpc, is seems unlikely the merger site 
should be outside either FOV.  Only cases where the kick velocity 
is extremely large, or the host galaxy's gravitational well is much
smaller than the Milky Way, allow mergers beyond this distance.

\section{Conclusions}

The high fraction of short GRBs with X-ray band afterglows,
and the potentially bright fluxes associated with them,
make the \textit{Swift} 
X-ray band an attractive wavelength to seek EM counterparts
to NS-NS and NS-BH mergers.  An imaging, wide field, soft X-ray band
detector with a fast response to TOO requests is required.
In the best case, the X-ray facility FOV would be at least
3 degrees wide to match the scale of LIGO/Virgo position uncertainty. 
However, given the full range of requirements, the \textit{Swift} satellite seems 
to be the strongest candidate facility which is currently 
in operation.
This paper discussed two possible search strategies,
that would likely be applicable under different sets of circumstances.

During the early years of advanced gravitational wave detectors,
around 2015-2018, Advanced LIGO and Advanced Virgo are likely
to operate at sensitivities less optimal than their design curves.
The sky-average range for NS-NS mergers will be perhaps 50 - 100 Mpc.  
Under these circumstances, the large ($\sim 100$ square degree) position
uncertainty associated with a LIGO/Virgo error box may be dramatically
reduced through the use of a galaxy catalog.  The \textit{Swift} observatory
could then search for an X-ray counterpart by imaging a few tens 
of galaxies over the course of a day, with exposures around
a kilosecond.  With this procedure, any on-axis afterglow
should be detectable, and so it should be possible to make
a detection, or else place limits on the beaming angle of 
short GRBs.  Moreover, while the XRT searches for an 
afterglow, the UVOT will simultaneously obtain data across
the band where kilonova emission is expected, imaging down to 
around magnitude 22, sufficiently deep to 
image a kilonova at 100 Mpc.  Kilonovae are expected to 
emit isotropically, and so could be observable even for 
off-axis merger events, suggesting that this observable could
accompany LIGO/Virgo triggers more often than afterglow 
emission.  During this period, the ``best guess'' estimate for
LIGO/Virgo detected mergers is only a few per year, so 
it seems plausible to follow-up every high significance trigger
in this manner.

As the GW detectors mature and reach their design sensitivity,
the galaxy catalog is likely to become a less useful tool.
This means that, instead of representing the error box with 
$\sim 30$ pointings, it will be necessary to use
hundreds of XRT tilings to cover the error box.  To complete
such an ambitious observing plan in a limited period of 
time requires sacrificing sensitivity, and the 100 s 
observations would be only barely deep enough to detect
on-axis afterglows, and would likely be unable to detect
kilonovae.  
On-axis afterglows are only expected to 
accompany $\leq 10\%$ of LIGO/Virgo detections, and 
the detection rate in this period could be 
$\geq 40$ per year.  Under these circumstances, it seems
unreasonable to expect an orbiting facility to 
follow-up every GW trigger.  
On the other hand, 
all-sky GRB monitors, most notably the \textit{Fermi} GBM,
continuously observe a large fraction of the sky, and
so effectively select which LIGO/Virgo triggers are 
most likely to have soft-band X-ray afterglows.   
Coincidences between a mature Advanced LIGO/Virgo network
and GBM should occur at the one per year level, an
estimate that comes from the observed GRB population,
and so is independent of large uncertainties in
population synthesis or the GRB jet opening angle \citep{coward_grb, holz2012}.  
A coincidence between GBM and the LIGO/Virgo network
will be an exciting event, and so obtaining the 
precise position, host galaxy, and red shift information
only obtainable through an afterglow observation 
will be well worth the effort.  Once the 
LIGO/Virgo network reaches its full design
sensitivity, with an average NS-NS inspiral 
horizon of around 200 Mpc, a sensible plan with
the \textit{Swift} X-ray observatory would be to 
only respond to triggers in coincidence
with a GRB observation.  
In this era ($\sim 2018$),
it is even possible that a fourth GW detector site will
be operational, and so reduce the sky area that needs to be 
searched.

\acknowledgements

We are grateful for fruitful discussions and feedback from
Lindy Blackburn,
David Burrows,
Thomas Dent, 
Brennan Hughey,
Susan Kassin,
Takanori Sakamoto, 
and Peter Shawhan.
JK is supported by an appointment to the NASA Postdoctoral 
Program at Goddard Space Flight Center, administered by Oak Ridge
Associated Universities through a contract with NASA.
This work made use of data supplied by the UK Swift Science Data Centre 
at the University of Leicester and data obtained from the High Energy Astrophysics Science Archive Research Center (HEASARC), provided by NASA's Goddard Space Flight Center.

\bibliographystyle{apj}
\bibliography{references}

\end{document}